\documentclass[prd,a4paper,twocolumn]{revtex4}
\usepackage{graphicx}
\usepackage{subfigure,amssymb}

\newcommand{\nc}{\newcommand}

\nc{\be}[1]{\begin{equation}\mbox{$\label{#1}$}}
\nc{\bea}[1]{\begin{eqnarray} \mbox{$\label{#1}$}}
\nc{\Section}[2]{\section{#2}\label{#1}}
\nc{\Bibitem}[1]{\bibitem{#1}}
\nc{\Label}[1]{\label{#1}}

\nc{\eea}{\end{eqnarray}}
\nc{\ee}{\end{equation}}

\nc{\bdm}{\begin{displaymath}}
\nc{\edm}{\end{displaymath}}
\nc{\dpsty}{\displaystyle}
\nc{\bc}{\begin{center}}
\nc{\ec}{\end{center}}
\nc{\ba}{\begin{array}}
\nc{\ea}{\end{array}}
\nc{\bab}{\begin{abstract}}
\nc{\eab}{\end{abstract}}
\nc{\btab}{\begin{tabular}}
\nc{\etab}{\end{tabular}}
\nc{\bit}{\begin{itemize}}
\nc{\eit}{\end{itemize}}
\nc{\ben}{\begin{enumerate}}
\nc{\een}{\end{enumerate}}
\nc{\bfig}{\begin{figure}}
\nc{\efig}{\end{figure}}

\nc{\arreq}{&\!=\!&}
\nc{\arrmi}{&\!-\!&}
\nc{\arrpl}{&\!+\!&}
\nc{\arrap}{&\!\!\!\approx\!\!\!&}
\nc{\non}{\nonumber}
\nc{\align}{\!\!\!\!\!\!\!\!&&}

\def\lsim{\; \raise0.3ex\hbox{$<$\kern-0.75em
      \raise-1.1ex\hbox{$\sim$}}\; }
\def\gsim{\; \raise0.3ex\hbox{$>$\kern-0.75em
      \raise-1.1ex\hbox{$\sim$}}\; }

\nc{\DOT}{\hspace{-0.08in}{\bf .}\hspace{0.1in}}
\nc{\Laada}{\hbox {$\sqcap$ \kern -1em $\sqcup$}}
\nc\loota{{\scriptstyle\sqcap\kern-0.55em\hbox{$\scriptstyle\sqcup$}}}
\nc\Loota{{\sqcap\kern-0.65em\hbox{$\sqcup$}}}
\nc\laada{\Loota}
\nc{\qed}{\hskip 3em \hbox{\BOX} \vskip 2ex}

\nc{\real}{{\rm I \! R}}
\nc{\Z}{{\sf Z \!\!\! Z}}
\nc{\complex}{{\rm C\!\!\! {\sf I}\,\,}}
\def\bigid{\leavevmode\hbox{\small1\kern-3.8pt\normalsize1}}
\def\id{\leavevmode\hbox{\small1\kern-3.3pt\normalsize1}}
\nc{\slask}{\!\!\!/}
\nc{\bis}{{\prime\prime}}
\nc{\pa}{\partial}
\nc{\na}{\nabla}
\nc{\ra}{\rangle}
\nc{\la}{\langle}
\nc{\goto}{\rightarrow}
\nc{\swap}{\leftrightarrow}

\nc{\EE}[1]{ \mbox{$\cdot10^{#1}$} }
\nc{\abs}[1]{\left|#1\right|}
\nc{\at}[2]{\left.#1\right|_{#2}}
\nc{\norm}[1]{\|#1\|}
\nc{\abscut}[2]{\Abs{#1}_{\scriptscriptstyle#2}}
\nc{\vek}[1]{{\rm\bf #1}}
\nc{\integral}[2]{\int\limits_{#1}^{#2}}
\nc{\inv}[1]{\frac{1}{#1}}
\nc{\dd}[2]{{{\partial #1}\over{\partial #2}}}
\nc{\ddd}[2]{{{{\partial}^2 #1}\over{\partial {#2}^2}}}
\nc{\dddd}[3]{{{{\partial}^2 #1}\over
    {\partial #2 \partial #3}}}
\nc{\dder}[2]{{{d #1}\over{d #2}}}
\nc{\ddder}[2]{{{d^2 #1}\over{d {#2}^2}}}
\nc{\dddder}[3]{{d^2 #1}\over
    {d #2 d #3}}
\nc{\dx}[1]{d\,^{#1}x}
\nc{\dy}[1]{d\,^{#1}y}
\nc{\dz}[1]{d\,^{#1}z}
\nc{\dl}[1]{\frac{d\,^{#1}l}{(2\pi)^{#1}}}
\nc{\dk}[1]{\frac{d\,^{#1}k}{(2\pi)^{#1}}}
\nc{\dq}[1]{\frac{d\,^{#1}q}{(2\pi)^{#1}}}

\nc{\bfT}{{\bf T }}

\nc{\cA}{{\cal A}}
\nc{\cB}{{\cal B}}
\nc{\cD}{{\cal D}}
\nc{\cE}{{\cal E}}
\nc{\cG}{{\cal G}}
\nc{\cH}{{\cal H}}
\nc{\cL}{{\cal L}}
\nc{\cO}{{\cal O}}
\nc{\cT}{{\cal T}}
\nc{\cN}{{\cal N}}
\nc{\cR}{{\cal R}}
\nc{\rvac}[1]{|{\cal O}#1\rangle}
\nc{\lvac}[1]{\langle{\cal O}#1|}
\nc{\rvacb}[1]{|{\cal O}_\beta #1\rangle}
\nc{\lvacb}[1]{\langle{\cal O}_\beta #1 |}
\nc{\bb}{\bar{\beta}}
\nc{\bt}{\tilde{\beta}}
\nc{\ctH}{\tilde{\cal H}}
\nc{\chH}{\hat{\cal H}}
\nc{\1}{\aa}
\nc{\2}{\"{a}}
\nc{\3}{\"{o}}
\nc{\4}{\AA}
\nc{\5}{\"{A}}
\nc{\6}{\"{O}}
\nc{\al}{\alpha}
\nc{\g}{\gamma}
\nc{\Del}{\Delta}
\nc{\e}{\textrm{e}}
\nc{\eps}{\epsilon}
\nc{\lam}{\lambda}
\nc{\Om}{\Omega}
\nc{\ve}{\varepsilon}
\nc{\mn}{{\mu\nu}}
\nc{\vp}{\varphi}

\nc{\rf}[1]{(\ref{#1})}
\nc{\nn}{\nonumber \\*}
\nc{\bfB}{\bf{B}}
\nc{\bfv}{\bf{v}}
\nc{\bfx}{\bf{x}}
\nc{\bfy}{\bf{y}}
\nc{\vx}{\vec{x}}
\nc{\vy}{\vec{y}}
\nc{\oB}{\overline{B}}
\nc{\oI}{\overline{I}}
\nc{\oR}{\overline{R}}
\nc{\rar}{\rightarrow}
\nc{\ti}{\times}
\nc{\slsh}{\hskip-5pt/}
\nc{\sm}{Standard~Model~}
\nc{\MP}{M_{\rm Pl}}
\nc{\mpl}{M_{\rm Pl}}
\nc{\tp}{t_{\rm Pl}}

\nc{\pmin}{p_{\rm min}}
\nc{\pmax}{p_{\rm max}}
\nc{\fo}{f_0}
\nc{\foi}{f_{0,i}\,}
\nc{\fop}{f_0^P}
\nc{\fou}{f_0^U}

\nc{\eff}{{\rm eff}}
\nc{\MT}{M_{\rm T}}
\nc{\ML}{M_{\rm L}}
\nc{\kk}{\vek{k}}
\nc{\pp}{{\rm p}}
\nc{\pt}{\partial_t}
\nc{\half}{{1\over 2}}
\nc{\w}{\omega}
\nc{\uhat}{\hat{U}_\w}

\nc{\etal}{\mbox{\it et al.}}
\nc{\ie}{{\it i.e. }}
\nc{\eg}{{\it e.g. }}
\nc{\trh}{T_{\rm RH}}
\nc{\ad}{{a'\over a}}
\nc{\bd}{{b'\over b}}
\nc{\Rd}{{R'\over R}}
\nc{\diag}{{\textrm{diag}}}
\nc{\mato}[1]{\tilde{#1}}
\nc{\sech}{\textrm{sech}}
\nc{\I}{\textrm{I}}
\nc{\II}{\textrm{II}}
\nc{\III}{\textrm{III}}
\nc{\vev}[1]{\langle #1 \rangle}
\nc{\hyp}{\,\; F_{1{\hskip -16pt}2}{\hskip 11pt}}
\nc{\brhom}{\overline{\rho}_M}
\nc{\brho}{\overline{\rho}}
\nc{\rhob}{\overline{\rho}}
\nc{\Pb}{\overline{P}}
\nc{\bH}{\overline{H}}
\nc{\ep}{{1+4\eps}}

\nc{\lcdm}{$\Lambda$CDM}

\def\smiley{\hbox{\large$\bigcirc$\hspace{-.80em}%
\raise.2ex\hbox{$\cdot\cdot$}\kern-.61em    
\lower.2ex\hbox{\scriptsize$\smile$}}\ }

\def\frowney{\hbox{\large$\bigcirc$\hspace{-.80em}%
\raise.2ex\hbox{$\cdot\cdot$}\kern-.635em
\lower.2ex\hbox{\scriptsize$\frown$}}\ }

\begin{document}

\title{Spherically symmetric solutions of modified field equations in $f(R)$ theories of gravity}

\author{T. Multam\"aki}
\email{tuomul@utu.fi}
\author{I. Vilja}
\email{vilja@utu.fi}
\affiliation{Department of Physics, University of Turku, FIN-20014 Turku, FINLAND}

\date{}

\begin{abstract}
Spherically symmetric static empty space solutions are studied in $f(R)$ theories of gravity.
We reduce the set of modified Einstein's equations to a single equation and show how
one can construct exact solutions in different $f(R)$ models. In particular, we show that for a 
large class models, including \eg the $f(R)=R-\mu^4/R$ model, the Schwarzschild-de Sitter 
metric is an exact solution of the field equations. The significance of these solutions is 
discussed in light of solar system constraints on $f(R)$ theories of gravity.
\end{abstract}
\maketitle

\section{Introduction}

The accelerating expansion of the universe has transformed our view of the universe from 
a matter filled cosmos to one dominated by dark energy.
Modern day cosmological observations are in contradiction with a matter dominated critical density 
universe whose expansion is decelerating and instead we find evidence for expansion that is accelerating.
Direct evidence supporting the cosmic acceleration comes from the supernovae observations \cite{snia}
and other observations, such as the cosmic microwave background \cite{cmb} and large scale structure 
\cite{lss}, provide more indirect evidence. Combining all of the observations, a cosmological 
concordance model has emerged: a critical density universe dominated by cold dark matter 
and cosmological constant -like dark energy. 

The most commonly considered candidate for dark energy is the cosmological constant 
(for a review see {\it e.g.} \cite{peebles}), but numerous alternative mechanisms for generating the 
cosmic acceleration have been considered. Very roughly, one can divide the different alternative
explanations of cosmic acceleration into two classes: those that include cosmic fluids with
exotic equations of state and those that modify gravity. In terms of the Friedmann equation
one can, again very roughly, consider the former to modify the right hand side of the equation,
the stress-energy tensor, and the latter the left hand side, the Einstein tensor.

Modifications of general relativity (GR) as a source of cosmic acceleration have been recently
considered in numerous works. One particular class of models that has drawn a significant amount
attention are the $f(R)$ gravity models
(see \eg \cite{turner,turner2,allemandi,meng,nojiri3,nojiri2,cappo1} and references therein).
These models are a particular class of higher derivative gravity theories that
include higher order curvature invariants as functions of the Ricci scalar. Such theories avoid the 
Ostrogradski's 
instability \cite{ostro} that can otherwise prove to be problematic for general higher derivative 
theories \cite{woodard,odintsov}.

A number of challenges have been identified in building phenomenologically viable models
of $f(R)$ gravity theories. Such possible obstacles include instabilities within matter \cite{dolgov},
outside matter \cite{soussa}, stability of the vacuum \cite{faraoni} and 
constraints arising from known properties of gravity
in our solar system (see {\it e.g.} \cite{chiba,confprobs,Clifton} and references therein). In addition,
identifying the specific functional form of $f(R)$ from cosmological observations is problematic
since the background expansion does not determine $f(R)$ uniquely \cite{Multamaki}.

In a number of works, the solar system constraints on $f(R)$ theories of gravity
are derived by first conformally transforming the theory to a scalar-tensor theory 
and then considering the Parameterized Post-Newtonian (PPN) limit \cite{damour,magnano}.
This procedure does not seem to be without controversy, however \cite{olmo, ppnok}. 
In this light, it is interesting to consider
solutions of the modified Einstein's equations of $f(R)$ theory. Armed with the metric, one can hope
to study orbital motion directly without resorting to conformal transformations.
As a first step in this direction, we consider vacuum solutions of the modified Einstein's equations in
this letter. We show how one can reduce the set of equations into a single equation that 
one can then utilize to construct explicit solutions. As an example we show that a large class of 
$f(R)$ models have the Schwarzschild-de Sitter metric as an exact solution
(see also \cite{cognola} for a related discussion). In addition, we construct 
other solutions corresponding to different metrics.

\section{$f(R)$ gravity formalism}

The action for $f(R)$ gravity is (see \eg \cite{Capozziello1})
\be{action}
S = \int{d^4x \sqrt{-g} \Big( f(R) + {\cal{L}}_{m} \Big)},
\ee
where we have set $8\pi G=1$. The field equations resulting from
this action in the metric approach, \ie assuming that the connection is the
Levi-Civita connection and variating with respect to  the metric $g_{\mu\nu}$, are 
\be{eequs}
G_{\mu\nu}\equiv R_{\mu\nu}-\frac 12 R g_{\mu\nu}=T^c_{\mu\nu}+T^m_{\mu\nu},
\ee
where the stress-energy tensor of the gravitational fluid is
\bea{tmunu}
T^c_{\mu\nu} & = & \frac{1}{F(R)}\Big\{\frac 12 g_{\mu\nu}\Big(f(R) - R\, F(R)
\Big)+\nonumber \\
& + & F(R)^{;\alpha\beta}\Big(g_{\alpha\mu}g_{\beta\nu}-g_{\mu\nu}
g_{\alpha\beta}\Big)\Big\}
\eea
with $F(R)\equiv d f(R)/dR$.

The standard minimally coupled stress-energy tensor $\tilde{T}_{\mu\nu}^m$, derived from 
the matter Lagrangian $\mathcal{L}_m$ 
in the action (\ref{action}), is related to $T_{\mu\nu}^m$ by
\be{matter}
T^m_{\mu\nu}=\tilde{T}^m_{\mu\nu}/F(R).
\ee
In empty space (vacuum), the equations of motion reduce to
\be{vacequ}
F(R)R_{\mu\nu}-\frac 12 f(R) g_{\mu\nu}-\nabla_\mu\nabla_\nu F(R)+
g_{\mu\nu}\Box F(R)=0.
\ee
Contracting these vacuum equations we obtain simply
\be{vaccontra}
F(R)R-2 f(R)+3\Box F(R)=0.
\ee
This equation is useful, because it allows us to express $f(R)$ in terms of its derivatives. If
$T^m_{\mu\nu}\ne 0$ there is an additional trace of stress-energy tensor $T^{m\,\mu}_\mu$ in the 
r.h.s. of  Eq. (\ref{vaccontra}).

\section{Spherically symmetric vacuum solutions}

We are interested in spherically symmetric, time independent solutions
of the empty space field equations. From properties of maximally symmetric sub-spaces
we know that the metric reads as
(in spherically symmetric coordinates)
\be{sphersym}
g_{\mu\nu}=\left(\begin{array}{cccc}
 s(t,r) & 0 & 0 & 0\\
0 & - p(t,r) & 0 & 0\\
0 & 0 & -r^2 & 0\\
0 & 0 & 0 & -r^2 \sin^2(\theta)
\end{array}
\right).
\ee
The (01)-component of the Einstein equations is satisfied if both $\dot p=0$ and 
$s(t,r)$ is separable with respect to its variables. This means that $\dot R=0$ and,
as in the well known case of the Schwarzschild metric, the time-dependence can be 
totally removed from the metric by redefinition of time. Here we make the same
simplifying assumption and thus we consider henceforth only time independent solutions
\ie $s=s(r)$ and $p=p(r)$.

The corresponding scalar curvature is
\be{sphersymR}
R=\frac{2}{r^2 p}\Big(1-p+(\frac{s'}{s}-\frac{p'}{p})(r-\frac{r^2}{4}
\frac{s'}{s})+\frac{r^2}{2}\frac{s''}{s}
\Big),
\ee
where we have defined $'\equiv d/dr$.

Using the contracted equation, Eq. (\ref{vaccontra}), 
the modified Einstein's equations become
\be{vacequ2}
F\, R_{\mu\nu}-\nabla_\mu\nabla_\nu F=\frac 14 g_{\mu\nu}(F\, R- \Box F).
\ee
Since the metric only depends on $r$, one can view Eq.\ (\ref{vacequ2}) as a 
set of differential
equations for $F(r),\ s(r)$ and $p(r)$. In this case both sides are diagonal and hence we 
have four equations. 
In addition, we have a consistency relation for $F(r)$,
\be{cons}
RF'-R'F+3(\Box F)'=0,
\ee
which arises by 
differentiating the contracted equation, Eq.\ (\ref{vaccontra}) with respect to $r$.
Any solution of Eq.\ (\ref{vacequ2}) must satisfy this relation in order to be 
also a solution of the original modified Einstein's equations, Eq.\ (\ref{vacequ}).

From Eq.\ (\ref{vacequ2}) it is obvious that the combination 
$A_{\mu}\equiv (F\, R_{\mu\mu}-\nabla_\mu\nabla_\mu F)/g_{\mu\mu}$ (with fixed indices)
is independent
of the index $\mu$ and therefore $A_\mu-A_\nu=0$ for all $\mu,\, \nu$. 
This allows us to write two equations:
\bea{Aequs}
2\frac{X'}{X}+r F'\frac{X'}{X}-2 r F'' & = & 0\label{aequ1}\\
-4 s+4 X-4 r s \frac{F'}{F}+2r^2s'\frac{F'}{F}+2rs\frac{X'}{X}- & & \nonumber\\
\ \ \ \ r^2s'\frac{X'}{X}+2r^2s'' & = & 0,\label{aequ2}
\eea
where we have defined $X(r)\equiv p(r) s(r)$. From Eq. (\ref{aequ1}) one can solve for $X'/X$ algebraically 
and substitute into Eq. (\ref{aequ2}) to obtain $X$:
\bea{xsolu}
X(r) & = & s(1+r\frac{F'}{F}-r^2\frac{F''}{2F+rF'}) + \nonumber\\
&  & \frac 12 r^2s'(r\frac{F''}{2F+rF'}-\frac{F'}{F})
-  \frac 12 r^2s''.
\eea
Consistency then requires that this form of $X(r)$ satisfies Eqs (\ref{aequ1}) and (\ref{aequ2}),
giving an equation relating $F$ and $s$. In addition, the modified Einstein's equations 
give four equations relating $F$ and $s$. However, all of the equations have a common
factor of the form:
\begin{widetext}
\bea{genequ}
s^{(3)}
& + & s''\frac{4{F}^2+4rFF'+r^2{F'}^2-3r^2FF''}{r F(2F+r F')}\nonumber\\
& - & s'\frac{8{F}^4+r^4{F'}^4+r^3F{F'}^2(4F'+rF'')+r^2{F}^2(6{F'}^2-3r^2{F''}^2+rF'(rF^{(3)}-2F''))+2r{F}^3(4F'+r(rF^3-F''))}{r^2{F}^2{(2F+rF')}^2}\nonumber\\
& + & 2s\frac{r^3{F'}^4+r^2F{F'}^2(3F'+rF'')+r^2{F}^2(-3r{F''}^2+F'(F''+rF^{(3)}))+{F}^3
(2r(2F''+rF^3)-4F')}{r^2{F}^2{(2F+rF')}^2}=0.
\eea
\end{widetext}
Therefore, any pair $s(r),\ F(r)$ satisfying this equation will be a solution of the modified
Einstein's equations. In addition, if Eq.\ (\ref{genequ}) is satisfied, also the consistency
relation, Eq.\ (\ref{cons}) is automatically satisfied. From $s$ and $F$ one can then calculate $R(r)$ 
and in principle construct the corresponding $f(R)$ by using Eq. (\ref{vaccontra}). 
Note that the resulting $f(R)$ is not unique due to the presence of an integration constant. 
In addition, a larger degeneracy can also exist, \eg for the SdS-solution discussed below,
$s(r)$ and $F(r)$ do not determine the $f(R)$ theory uniquely, even when discounting the integration
constant.

\subsection{Solutions with constant curvature}

Looking for constant curvature solutions, $R=R_0$, the  field equations reduce to
\bea{constcurv}
sp'+ps' & = & 0,\\
1-p+\frac r2(\frac{p'}{p}+\frac{s'}{s})(\frac{r}{2}\frac{s'}{s}-1)
-\frac{r^2}{2}\frac{s''}{s} & = & 0,
\eea
which are straightforwardly solvable:
\bea{ccsols}
p(r) & = & \frac{c_0}{s(r)}\nonumber\\
s(r) & = & c_0+\frac{c_1}{r}+c_2 r^2,
\eea
where $c_i$ are integration constants. For conventional definitions of space and time
we require $c_0>0$.
The scalar curvature Eq. (\ref{sphersymR}) for this solution
is $R=12 c_2/c_0$. Redefining the time coordinate, $t\goto t/\sqrt c_0$ 
with $c_1 \goto c_1/\sqrt c_0$ and $c_2 \goto c_2/\sqrt c_0$, we can always choose
$c_0=1$.

The Schwarzschild solution in the presence of a cosmological constant, Schwarzschild-de Sitter -spacetime (SdS) 
arising using $f(R)= R+ \Lambda$, has the form
\be{schwlam}
g_{\mu\nu}=\left(\begin{array}{cccc}
A(r)& 0 & 0 & 0\\
0 & -1/A(r) & 0 & 0\\
0 & 0 & -r^2 & 0\\
0 & 0 & 0 & -r^2 \sin^2(\theta)
\end{array}
\right),
\ee
where $A(r)\equiv 1-2M/r-\Lambda r^2/3$ and $\Lambda$ is the cosmological constant. 
The scalar curvature in this case is a constant, $R=-4\Lambda$. 
For a finite mass distribution, the parameter $M$ can in this case be identified as total material mass  
\be{cm}
M_c=\int^{r_m} d^3x\, T^{m\,0}_0.
\ee
Note that this integral also contains a part of gravitational energy inside the radius $r_m$ of the 
mass distribution \cite{weinberg}, while the total energy within the same radius also includes vacuum energy, $E_{tot}=\int d^3x\, (T^{m\,0}_0+ \Lambda)$.

Comparing the solution, Eq. (\ref{ccsols}), to the SdS metric, Eq.\ (\ref{schwlam}),
one sees that the two metrics are identical with $c_1=-2M,\ c_2=-\Lambda/3$.
From Eq.\ (\ref{vaccontra}) it is clear that this metric is a solution
for any form of $f(R)$ for which there exists a constant (real) $R_0$ such that 
$R_0 f'(R_0)-2 f(R_0)=0$. In other words, the SdS metric is an exact solution
for a set of functions $f(R)$ that satisfy $R_0 f'(R_0)-2f(R_0)=0$ such that $R_0$ is real.
For example, for the $f(R)=R-\mu^4/R$ model, it is easy to see that the
SdS metric is a solution when $R_0^2=3\mu^4=144 c_2^2$, {\it i.e.}
$c_2=\sqrt{3\mu^4/144}$. The same exact result also holds for the
other commonly considered model $f(R)=R-\mu^4/R+\eps R^2$.

The physical interpretation of the parameters $M$ and $\Lambda$ are not as 
straightforward as in the case of general relativity.
The naive identification $M=M_c$, is problematic and a more careful analysis is required.
In the presence on spherically finite symmetric mass distribution the empty space solution 
we are studying needs to be matched to the solution 
valid inside the mass distribution at 
$r=r_m$. Since the field equations are in general higher order
differential equations than in GR, more integration constants need to be determined. 
In particular, this means that values of the metric components inside the mass distribution
depend explicitly on the details of the mass distribution, making matching with the outside solution
non-unique. This can be explicitly seen
by studying {\it e.g.} solutions of spherical shells of different thicknesses: the boundary values at
$r_m$ depend explicitly on the thickness of the shell.
This also demonstrates that the
Birkhoff theorem is no longer valid since the external solution depends on the internal mass distribution. 
This is a general property of all (empty space) solutions of $f(R)$ theories whenever $F$ differs from a 
constant. If we define the central mass by
the gravitational effect it gives rise to the external space, the parameter $M$ becomes defined as
the central mass, but it does not coincide with (\ref{cm}).

\subsection{General solutions with $p(r)s(r)=const.$}

From Eq.\ (\ref{aequ1}), it is clear that when $X=const.\equiv X_0$, $F''(r)=0$ and hence
$F(r)=A\, r+B$. Eq.\ (\ref{genequ}) can now be solved, giving
\begin{widetext}
\bea{x1solu}
s(r) & = & 
X_0 + \frac{Ac_1}{2B^2} - \frac{c_1}{3Br} - r
  \frac{{A}\left(B^2 X_0 + Ac_1 \right) }
   {{{B}}^3}\nonumber\\
& + & r^2\frac{
     3A^2B^2X_0 + 2B^4c_2
    +   2A^2\left( B^2X_0 + Ac_1 \right) \ln | B/r + A| }{2B^4},
\eea
\end{widetext}
where $c_i$ are constants. Requiring a SdS-type solution, we
must choose $X_0=1$, which then sets $A=0$ and we may write also $c_1=6BM$, reducing  
Eq.\ (\ref{x1solu})
to $s(r)=1-2M/r+c_2 r^2$ with constant curvature. If we were to choose $c_1=0$ instead, 
the mass term would be absent. It is, however, unclear whether these solutions correspond to 
maximally  symmetric (spatial) spaces or to some other type of spherically symmetric 
(but non-singular) cases.

Requiring that the SdS-type metric is a solution is hence equivalent to requiring constant
scalar curvature and the conclusions of the previous section apply.

As in the constant curvature case, time can always be rescaled so that
$X_0=1$. (Alternatively we can choose the time scaling of $s(r)$ so that 
$X_0 + Ac_1/2B^2=1$, which generally leads to $X_0\ne 1$. For examples, see section \ref{ES}.) 
Taking $X_0=1$ we find that for small values of the radial coordinate $r$, the leading terms of the 
general solution (\ref{x1solu}) read $s(r) \sim 1 + Ac_1/2B^2 - c_1/3Br$ where
we again identify $c_1=6BM$, leading to the correct form of Newtonian potential. However, additional corrections
to the geodetic motion appear because in general $Ac_1/2B^2\ne 0$, giving raise
to additional parameter constraints. In the large $r$ limit the leading contribution comes from 
the $r^2$ -term. It should be noted, however, that $s(r)$ may have large finite zeros like in the 
SdS solution, making the limit $r\goto \infty$ physically uninteresting.

\subsection{Asymptotic solutions}

In order to have a better handle on the question of uniqueness of the solutions, 
we consider asymptotic solutions for which $s(r)\rightarrow 1$ at large $r$, 
mimicking the standard Schwarzschild-solution. Our {\it Ans\"atze} are
\bea{ansa}
s(r) & = & 1-2M/r\nonumber\\
F(r) & = & F_0 r^n\nonumber.
\eea
Inserting these into the modified Einstein's equations, by requiring that the highest
order term vanishes one finds that $n=0$ or $F(r)=const.$ is the leading term in $F$.
Hence there are no new solutions that tend to constant scalar curvature in the large 
$r$ limit along with $s(r)$, suggesting that any new solutions will be radically 
different from  the Schwarzschild(-de Sitter) solution.

\subsection{Exact solutions}\label{ES}

We have also found a number of exact solutions to Eq.\ (\ref{genequ}), and by considering different
Ans\"atze for $s(r)$ or $F(r)$ one can easily find and construct more solutions. A number of 
interesting solutions along with the corresponding forms of $f(R)$ are:
\begin{widetext}
\be{solulist}
\begin{array}{c|c|c|c|c|c}
& s(r) & F(r) & X(r) & f(R) & R\\
\hline
I & 1-\frac{2M}{r}-\frac 13\Lambda r^2 & 1-\frac{1}{3M}r & 2 & R\pm \frac{2}{3M}\sqrt{-R-2\Lambda}+\Lambda & -\frac{1}{r^2}-2\Lambda \\
II & 1-\frac 13\Lambda r^2 & F_0 r & 2 & \pm 2\sqrt{-R-2\Lambda} & -\frac{1}{r^2}-2\Lambda\\
III & s_0 r^m & F_0 r^n,\, n=2\frac{m(m-1)}{m-2} & -2\frac{n^4-n^3-4n-2}{(n+2)^2}s_0r^m & 
\frac{2F_0}{2-n}\left(\frac{3n(2-n)}{n^2-2n-2}\right)^{n/2}R^{1-n} & \frac{3(2-n)n}{n^2-2n-2}
r^{-2}.
\end{array}
\ee
\end{widetext}

These solutions, in particular $I$ and $II$, could be considered as suitable asymptotic limit, either $R\goto 0$ or $R \goto -\infty$, of a more general $f(R)$ having linear $f(R) \propto R$ term.

\section{Discussion and conclusions}

We have seen that the set of Einstein's equations reduce to a single non-linear
differential equation relating $s(r)$ and $F(r)$. In spite of the complicated form,
a number of solutions is straightforwardly found. The applicability of the general solutions
could probably be tested \eg by exploiting parameter constraints appearing from solar system 
and comparing these with those arising from cosmology.

Considering the SdS-solution that is present in a large class of models, 
\eg the $R-\mu^4/R$ model with $R_0^2=3\mu^4$ or $\Lambda^2=3\mu^4/16$, 
the parameter $\Lambda$ can be
constrained by a number of different observations in the solar system.
Such are the gravitational redshift measurements, gravitational time
delay measurements by the Cassini spacecraft and the perihelion shift of Mercury
(see {\it e.g.} \cite{kagramanova}). 
The tightest constraint arises from the perihelion shift of Mercury, 
for which it is found \cite{kagramanova} that $|\Lambda|<10^{-41}m^{-2}$
(the cosmologically observed value is roughly $10^{-52}m^{-2}$). 
The solar system observations are hence not able to effectively constrain such a metric
compared to the cosmologically relevant values. For the other solutions we have found,
solar system observations are likely to be more efficient (\eg for $X\neq 1$).

In terms of the equivalent scalar-tensor theory, the SdS-solution corresponds
to a constant field solution. It is straightforward to see that the effective scalar 
mass, $m^2=V''(\phi)$, is positive when $f'(R_0)/f''(R_0)-2f(R_0)/f'(R_0)>0$.
This is equivalent to requiring that the vacuum state is stable with respect to small 
perturbations \cite{faraoni}. However, it is not at all clear what,
if any, role the effective scalar plays since the metric solution is independent
of the scalar mass. This question will be addressed in further work.

Important questions in addressing the validity of the metric solutions presented here
are the stability and uniqueness of the solution. For example, in order for the SdS metric to 
be physically relevant, it must be stable with respect to small perturbations \ie instead 
of a test mass, one needs to consider the effect of a massive body on the metric. 
Uniqueness of $f(R)$ is also an interesting question assessing the physical relevance of a 
solution. Asymptotic considerations indicate that Schwarzschild-type metrics lead to 
constant curvature suggesting that new solutions will deviate strongly from the standard 
Schwarzschild metric. 

By considering the boundary conditions of the general SdS-solution, we have noted that 
the Birkhoff theorem is not valid for non-trivial $f(R)$ theories as the solutions around 
spherically symmetric mass distribution 
depends on the shape of the distribution. Hence the straightforward Schwarzschildian relation 
between the gravitational effect and total energy of a finite mass distribution has been broken. 
By requiring that it holds, we are restricted to some class of special mass distributions. Also 
we are lead to ask which distributions are physically relevant and what kind of distributions are 
likely to form from collapsing matter. These most interesting but also technically extremely 
tricky questions certainly require further examination because they may offer additional
constraints on allowed $f(R)$ models.

Our results show that in addition to the SdS metric, $f(R)$ theories typically also have
new different solutions. Although further work is needed to determine their physical 
relevance, they offer an interesting new avenue of research that can guide us in assessing
the significance of $f(R)$ theories of gravity as a possible solution to the dark energy problem.

\acknowledgments
TM would like to thank K. Kainulainen and D. Sunhede for fruitful discussions. TM is
supported by the Academy of Finland.



\end{document}